\documentclass[useAMS]{mn2e}
\title[]{Time Dependence of Advection Dominated Accretion Flow with a Toroidal Magnetic Field}
\author[A. Khesali and K. Faghei]
{Alireza Khesali \thanks{E-mail: khesali@umz.ac.ir}
and Kazem Faghei \thanks{E-mail: faghei@umz.ac.ir} \\
Department of Physics, Mazandaran University, Babolsar,
Iran }
\begin{document}

\date{}

%\pagerange{\pageref{firstpage}--\pageref{lastpage}} \pubyear{0000}

\maketitle

\label{firstpage}

\begin{abstract}
The present study examines self-similarity evolution of advection
dominated accretion flow (ADAF) in the presence of a toroidal
magnetic field. In this research, it was assumed that the angular
momentum transport is due to viscous turbulence and
$\alpha$-prescription was used for kinematics coefficient of
viscosity. The flow does not have a good cooling efficiency and
so, a fraction of energy accretes with matter on central object.
The effect of a toroidal magnetic field on such systems in a
dynamical behavior was investigated. In order to solve the
integrated equations which govern the dynamical behavior of the
accretion flow, self-similar solution was used. The solution
provides some insights into the dynamics of quasi-spherical
accretion flow and avoids many of the strictures of the steady
self-similar solutions. The solutions show that the behavior of
physical quantities in a dynamical ADAF are different from steady
accretion flow and a disk with polytropic approach. The effect of
the toroidal magnetic field is considered with additional
variable $\beta[=p_{mag}/p_{gas}]$, where $p_{mag}$ and $p_{gas}$
are the magnetic and gas pressure, respectively. Also to consider
the effect of advection in these systems, the advection parameter
$f$ was introduced that stands for a fraction of energy that
accretes by matter to the central object. The solution indicates
a transonic point in the accretion flow for all selected amounts
of $f$ and $\beta$. Also, by adding strength of the magnetic
field and the degree of advection, the radial-thickness of the
disk decreased and the disk compressed. The model implies that
the flow has differential rotation and is sub-Keplerian at small
radii and is super-Keplerian in large radii and that different
result was obtained using a polytropic accretion flow. The
obtained $\beta$ parameter was used a function of position that
increases by increasing radii. Also, The behavior of ADAF in a
large toroidal magnetic field implies that different result was
obtained using steady self-similar models in large magnetic field.
\end{abstract}


\section{Introduction}
During recent years one type of accretion disks has been studied,
in which it is assumed that the energy released through viscous
processes in the disk may be trapped within the accreting gas.
This kind of flow is known as advection-dominated accretion flow
(ADAF). The basic ideas of such ADAF models have been developed
by a number of researchers (e.g., Ichimaru 1977; Rees et al.
1982; Narayan \& Yi 1994, 1995; Abramowicz et al. 1995; Ogilvie
1998; Akizuki \& Fukue 2006; hereafter AF06).

It is thought that accretion disks, whether in star-forming
regions, in X-ray binaries, in cataclysmic variables, or in the
centers of active galactic nuclei, are likely to be threaded by
magnetic fields. Consequently, the role of magnetic fields on
ADAF has been analyzed in detail by a number of investigators
(Bisnovatyi-kogan \& Lovelace 2001; Shadmehri 2004; AF06;
Ghanbari et al. 2007, Abbassi et al. 2008). The existence of the
toroidal magnetic fields have been proven in the outer regions of
YSO discs (Greaves et al. 1997; Aitken et al. 1993; Wright et al.
1993) and in the Galactic center (Novak et al. 2003; Chuss et al.
2003). Thus, considering the accretion disks with a toroidal
magnetic field have been studied by several authors (AF06;
Begelman \& Pringle 2007; abbassi et al. 2008; Khesali \& Faghei
2008 and references within; hereafter KF08). KF08 considered
dynamic behavior of a polytropic accretion flow in presence of a
toroidal magnetic field. In a dynamic approach they showed the
radial behavior of the physical quantities were different with
results achieved by those who considered the accretion flow in a
steady self-similar methods (Shadmehri 2004; AF06; Ghanbari et al
2007; Abbassi et al. 2008). For example, KF08 presented that
ratio of the magnetic pressure to the gas pressure is not
constant and varies by radius. The results of KF08 were assembled
on polytropic equation that implies the accreting gas has a good
cooling efficiency, while results of some authors have shown that
the behavior of physical quantities are very sensible to fraction
of the energy that traps within the accreting gas (AF06). So,in
the present study it is intended to investigate dynamic behavior
of an ADAF in presence of a toroidal magnetic field. The paper is
organized as follows. In section 2, the general problem of
constructing a model for quasi-spherical magnetized advection
dominated accretion flow will be defined. In section 3,
self-similar method for solving the integrated equations which
govern the dynamic behavior of the accreting gas was utilized.
The summary of the model will appear in section 4.
\section{Basic equations}
In this section, we derive the basic equations which describe the
physics of accretion flow with a toroidal magnetic field. We use
the spherical coordinates $(r, \theta, \phi)$ centred on the
accreting object and make the following standard assumptions:
\begin{itemize}
  \item [(i)] The accreting gas is a highly ionized gas with infinitive conductivity;
  \item [(ii)] The magnetic field has only an azimuthal component;
  \item [(iii)]The gravitational force on a fluid element is characterized by the Newtonian potential
             of a point mass, $\Psi=-GM_{*}/r$,
  with $G$ representing the gravitational constant and $M_{*}$ standing for the mass of the central star;
  \item [(iv)] The equations written in spherical coordinates are considered in the equatorial
  plane $\theta=\pi/2$ and  terms with any $\theta$ and $\varphi$ dependence are neglected, hence
  all quantities will be expressed in terms of spherical radius $r$ and time $t$;
  \item [(v)] For simplicity, the self-gravity and general relativistic effects have been neglected.
\end{itemize}
Under the assumptions and the approximation of quasi-spherical symmetry and the ideal
magnetohydrodynamics treatment, the dynamics of a magnetized accretion flow is described by the following equations:\\
the continuity equation
\begin{equation}\label{a1}
\frac{\partial\rho}{\partial t}
+\frac{1}{r^{2}}\frac{\partial}{\partial r}(r^{2}\rho v_{r})=0,
\end{equation}
the radial force equation
\begin{equation}\label{a2}
\frac{\partial v_{r}}{\partial t}+v_{r}\frac{\partial
v_{r}}{\partial r}
  +\frac{1}{\rho}\frac{\partial p}{\partial r}
  +\frac{GM_{*}}{r^{2}}
   =r\Omega^{2}-\frac{B_{\varphi}}
     {4 \pi r\rho}\frac{\partial }{\partial r}(rB_{\varphi}),
\end{equation}
the azimuthal force equation
\begin{equation}\label{a3}
\rho\left[\frac{\partial}{\partial
t}(r^{2}\Omega)+v_{r}\frac{\partial}{\partial r}(r^{2}\Omega)\right]=
  \frac{1}{r^{2}}\frac{\partial}{\partial r}\left[\nu\rho r^{4}\frac{\partial \Omega}{\partial r}\right],
\end{equation}
the energy equation
\begin{eqnarray}
 \nonumber \frac{1}{\gamma-1}\left[\frac{\partial p}{\partial t}+ v_r\frac{\partial p}{\partial r}\right]+
\frac{\gamma}{\gamma-1}\frac{p}{r^2}\frac{\partial}{\partial
r}\left( r^2 v_r\right)= \\
 f \nu\rho r^2
\left(\frac{\partial\Omega}{\partial r}\right)^2
\end{eqnarray}
and the field freezing equation
\begin{eqnarray}\label{a5}
\frac{\partial B_{\varphi}}{\partial t} +\frac{1}{r}\frac{\partial
}{\partial r}(rv_{r}B_{\varphi})=0,
\end{eqnarray}
Here $\rho$ is the density, $v_r$ the radial velocity, $\Omega$
the angular velocity, $M_*$ the mass of the central object, $p$
the gas pressure, $B_{\varphi}$ the toroidal component of
magnetic field, $\nu$ the kinematic viscosity coefficient and it
is given, as in Narayan \& Yi (1995), by an $\alpha$-model

\begin{equation}
   \nu = \alpha \frac{p_{gas}}{\rho\Omega_{K}}
\end{equation}
where $\Omega_K=({GM_*}/{r^3})^{1/2}$ is the Keplerian angular
velocity. The parameters $\gamma$ and $\alpha$ are assumed to be
constant and $f$ measures the degree to which the flow is
advection-dominated (Narayan \& Yi 1994), and is assumed to be
constant.
\section{Self-similar solutions}
\subsection{Analysis}
Self-similar models have proved very useful in astrophysics
because the similarity assumption reduces the complexity of the
partial differential equations. Even greater simplification is
achieved in the case of spherical symmetry since the governing
equations then reduce to comparatively simple ordinary
differential equations. We introduce a similarity variable $\eta$
and assume that each physical quantity is given by the following
form:
\begin{equation}\label{a11}
r=r_0(t)\eta
\end{equation}
\begin{equation}\label{a12}
\rho(r,t)=\rho_0(t)R(\eta)
\end{equation}
\begin{equation}\label{a122}
p(r,t)=p_0(t)P(\eta)
\end{equation}
\begin{equation}\label{a13}
v_{r}(r,t)=v_0(t)V(\eta)
\end{equation}
\begin{equation}\label{a14}
\Omega(r,t)=\Omega_0(t)\omega(\eta)
\end{equation}
\begin{equation}\label{a15}
B_{\varphi}(r,t)=b_0(t)B(\eta).
\end{equation}
By assuming power-law time dependent of $r_0(t)=at^n$, where
$n=2/3$, we find the following relations:

\begin{equation}
r_0(t)=at^{2/3}
\end{equation}
\begin{equation}
p_0(t)/\rho_0(t)=\frac{GM_*}{a}t^{-2/3}
\end{equation}
\begin{equation}
v_0(t)=\sqrt{\frac{GM_*}{a}}t^{-1/3}
\end{equation}
\begin{equation}
\Omega_0(t)=\sqrt{\frac{GM_*}{a^3}}t^{-1}
\end{equation}
\begin{equation}
b_0^2(t)/8\pi \rho_0(t)=\frac{GM_*}{a}t^{-2/3}.
\end{equation}
The above results imply that $p_0(t)$ and $b_0(t)$ are dependent
on timely behavior of $\rho_0(t)$. So, for specifying time
dependent of $\rho_0(t)$, and then $p_0(t)$ and $b_0(t)$, we
introduce the mass accretion rate $\dot{M}$

\begin{equation}
  \dot{M}=-4 \pi r^2 \rho v_{r}.
\end{equation}
Similar to equations (7)-(12) for the mass accretion rate we can write
\begin{equation}
  \dot{M}(r,t)=\dot{M}_0(t)\dot{m}(\eta).
\end{equation}
Under transformations of equations (7), (8) and (10), equation (19) becomes
\begin{equation}
  \dot{M}(r,t)= \left[r_0^2(t) \rho_0(t) v_0(t)\right]\times\left[ -4 \pi\eta^2 R(\eta) V(\eta)\right]
\end{equation}
in which implies
\begin{equation}
  \dot{M}_0(t)=r_0^2(t) \rho_0(t) v_0(t)
\end{equation}
\begin{equation}
  \dot{m}(\eta)=-4 \pi\eta^2 R(\eta) V(\eta).
\end{equation}
Now, we consider a set of solutions that $\dot{M}_0(t)$ is a constant (KF08), thus we can write
\begin{equation}
\rho_0(t)=(\dot{M}_0/\sqrt{GM_*a^3})t^{-1}
\end{equation}
that implies
\begin{equation}
p_0(t)=(\dot{M}_0\sqrt{GM_*/a^5})t^{-5/3}
\end{equation}
and
\begin{equation}
b^2_0(t)/8\pi=(\dot{M}_0\sqrt{GM_*/a^5})t^{-5/3}.
\end{equation}
Substituting equations (6)-(12) and (13)-(17) into the basic equations (1)-(6), the similarity equations are obtained as
\begin{equation}\label{a20}
-R+\left(V-\frac{2\eta}{3}\right)\frac{dR}{d\eta}+\frac{R}{\eta^2}
\frac{d}{d\eta}\left(\eta^2V\right)=0,
\end{equation}
\begin{eqnarray}\label{a21}
\nonumber -\frac{V}{3}+\left(V-\frac{2\eta}{3}\right)\frac{dV}{d\eta}
+\frac{1}{R}\frac{dP}{d\eta}+\frac{1}{\eta^{2}}=
~~~~~~~~~~~~~\\  \eta\omega^{2}-\frac{2 B}{\eta R}
\frac{d\left(\eta B\right)}{d\eta},
\end{eqnarray}
\begin{eqnarray}\label{a22}
\nonumber R\left[\frac{1}{3}\left(\eta^{2}\omega\right)+\left(V-\frac{2\eta}{3}
\right)\frac{d}{d\eta}\left(\eta^{2}\omega\right)
\right]=~~~~~~~~~~~~~~~\\\frac{\alpha}{\eta^{2}}\frac{d}{d\eta}
\left[P\eta^{11/2}\frac{d\omega}{d\eta}\right],
\end{eqnarray}

\begin{eqnarray}
\nonumber\frac{1}{\gamma-1}\left[-\frac{5}{4}P+\left(V-\frac{2\eta}{3}\right)\frac{dP}{d\eta}\right]+
\frac{\gamma}{\gamma-1}\frac{P}{\eta^2}\frac{d}{d\eta}\left(\eta^2V
\right)\\
= \alpha f P \eta^{7/2}\left(\frac{d\omega}{d\eta}\right)^2,
\end{eqnarray}

\begin{equation}\label{a23}
-\frac{5}{4}B+\left(V-\frac{2\eta}{3}
\right)\frac{dB}{d\eta}+
\frac{B}{\eta}\frac{d}{d\eta}\left(\eta V\right)=0.
\end{equation}

To investigate existence of transonic point, the square of the
sound velocity is introduced that subsequently can be expressed as
\begin{equation}\label{a35}
v_{s}^2 \equiv \frac{p}{\rho}=\frac{GM_*}{a}\frac{P}{R}~t^{-2/3}
\end{equation}
Here, $S=\left(P/R\right)^{1/2}$ the \emph{sound velocity}
in self-similar flow, which is rescaled in the course of time.
The \emph{Mach number} referred to the reference frame is defined as (Fukue 1984;
Gaffet \& Fukue 1983)
\begin{equation}\label{a36}
    \mu\equiv\frac{v_{r}-v_{F}}{v_{s}}=\frac{V-n\eta}{S}
\end{equation}
where
\begin{equation}\label{a37}
    v_{F}=\frac{dr}{dt}=n\frac{r}{t}
\end{equation}
is the velocity of the reference frame which is moving outward as
time goes by. The Mach number introduced so far, represents the
\emph{instantaneous} and \emph{local} Mach number of the unsteady
self-similar flow. We will consider transonic points of accretion
flow in next subsection.

In order to consider the strength of the magnetic field in the
plasma, the $\beta$ parameter is introduced that is ratio of the
magnetic to the gas pressures
\begin{equation}
\beta(r,t)=\frac{B^2_{\varphi}(r,t)/8\pi}{ p(r,t)}=\frac{B^2(\eta)}{P(\eta)}.
\end{equation}

In completing this section, we also summarize the main results
here. Solving equations (1), (10), (11), and (19) under
transformations (12)-(15) in non-magnetically state, makes it
clear that time behavior of physical quantities in the non-
magnetically and the magnetically disk are the same. This result
is one of the strictures of time-dependent self-similar solution.
on the other hand, the fact that timely- dependent behavior of
the magnetic and gas pressures becomes same is one of limits the
self-similarity solution. On the other hand, the physical
quantities with a same physical dimension have similar behaviors in self similar solution.
\subsection{Asymptotic behavior}
In this subsection, the asymptotic behavior of the equations (22), (26)-(30), and (34) at $\eta\rightarrow0$ and $\gamma < 5/3$ is investigated. the asymptotic solutions are given by
\begin{equation}
    R(\eta) \sim R_{0}\eta^{-3/2}
\end{equation}
\begin{equation}
    P(\eta)\sim P_{0}\eta^{-5/2}
\end{equation}
\begin{equation}
    V(\eta)\sim V_{0}\eta^{-1/2}
\end{equation}
\begin{equation}
    \omega(\eta)\sim \omega_{0}\eta^{-3/2}
\end{equation}
\begin{equation}
    B(\eta)\sim B_{0}\eta^{-1/2}
\end{equation}
\begin{equation}
    \dot{m}(\eta)\sim -4\pi R_{0} V_{0}
\end{equation}
\begin{equation}
    \beta(\eta)\sim ({B^2_{0}}/{P_0})\eta^{3/2}
\end{equation}
in which

\begin{equation}
    R_0 =-\frac{3}{8\pi}\alpha f \dot{m}_{in} \left(\frac{\gamma-1}{\gamma-5/3}\right)\left(\frac{g_1}{g_3}\right)
\end{equation}
\begin{equation}
    P_0=\frac{\dot{m}_{in}}{6\pi\alpha}
\end{equation}
\begin{equation}
   V_0 = \frac{2}{3\alpha f}\left(\frac{\gamma-5/3}{\gamma-1}\right)\left(\frac{g_3}{g_1}\right)
\end{equation}
\begin{equation}
    \omega_0=-\frac{2}{3\alpha f}\left(\frac{\gamma-5/3}{\gamma-1}\right)\left(\frac{g_3}{g_1}\right)^{1/2}
\end{equation}
\begin{equation}
    B^2_0=\beta_{0}\frac{\dot{m}_{in}}{6\pi\alpha}
\end{equation}

where

\begin{equation}
    \frac{1}{g_1}=1-\frac{5f}{2}\left(\frac{\gamma-1}{\gamma-5/3}\right)
\end{equation}
\begin{equation}
    g_2=\frac{3}{2}\alpha f \left(\frac{\gamma-1}{\gamma-5/3}\right)
\end{equation}
\begin{equation}
    g_3=-1+\sqrt{1+2 g^2_1 g^2_2}
\end{equation}
\begin{equation}
    \beta_{0}=\beta_{in}/\eta^{3/2}_{in}.
\end{equation}

The achieved results for asymptotic behavior of physical
quantities show that the physical quantities of accretion flow
are very sensible to parameters of $\alpha$, $\gamma$, $f$,
$\beta_{in}$, and $\dot{m}_{in}$.  The $\beta_{in}$ and
$\dot{m}_{in}$ are amounts of $\beta$ and $\dot{m}$ at
$\eta_{in}$ that $\eta_{in}$ is a point near of the center. The
affects of the viscous parameter $\alpha$ and the advection
parameter $f$ on accretion flow are plotted in figure 1. The
angular velocity profiles indicate that by increasing the viscous
parameter $\alpha$, the angular velocity of accretion flow
decreases, because we increase the viscous torque by increasing
parameter $\alpha$. Also increasing the advection parameter $f$
decreases the angular velocity that is qualitatively consistent
with AF06. Figure 1 shows the radial infall velocity increases by
adding $\alpha$ and $f$ that are similar to the results of AF06
and KF08. Also the density profiles represent density decreases
by adding $f$ and $\alpha$.

\subsection{Numerical solutions}
If the value of $\eta_{in}$ is guessed, that is a point very near
to the center, the equations can be integrated from this point to
the outward through the use of the above expansion. Examples of
such solutions are presented in figures 2, 3, and 4. The profiles
in figure 2 are plotted for different $\beta_{in}$, the profiles
in figure 3 are plotted for different $f$ and in figure 4
transonic behavior of the accreting gas for different amount of
$f$ and $\beta_{in}$ is considered. The delineated quantities
($Log (\eta^{3/2}R)$, $Log(-\eta^{1/2}V)$, ...) in figures 2, 3,
and 4 are constant in steady self-similar solutions (Narayan \& Yi
1994; Narayan \& Yi 1995; Shadmehri 2004; AF06; Ghanbari et. al.
2007; Abbassi et al. 2008), while here, they vary by position.

Figure 2 informs us that density and the radial thickness of disk
decreases by adding strength of the toroidal magnetic field,
these results are well consistent with KF08. Also, by decreasing
amount of magnetic field, the behavior of density becomes similar
to non-magnetic case (Ogilvie 1999). The behavior of the gas
pressure in KF08 had polytropic behavior and this selection
caused the gas pressure follow the density behavior, while here
we see behavior of the gas pressure does not follow the density
behavior. Also, by adding the $\beta$ parameter, the radial
infall velocity increases; such property is qualitatively
consistent with AF07 and KF08. This is due to the magnetic
tension terms, which dominate the magnetic pressure term in the
radial momentum equation that assist the radial infall motion.
The profiles of the angular velocity imply that the disk is
sub-Keplerican in inner part of the disk and is super-Keplerian
in outer part of it, while in polytropic accreting flow
(KF08) and non-magnetic accretion flow (Ogilvie 1999) the angular
velocity is sub-Keplerian in all radii (KF08). Similar to the
results KF08 the $\beta$ parameter, the ratio of the magnetic
pressure to the gas pressure, is a function of position and
arises from inner to outer that the result is well consistent with
observational evidence obtained by some authors (Aitken et al.
1993; Wright et al. 1993; Greaves et al. 1997). While the $\beta$
parameter in steady self-similar solution  becomes constant at all
radii (AF06) that is one of restriction of steady self-similar
solution. Figure 3 is plotted for different amounts of
the advection parameter $f$. The advection parameter $f$ has
slight effect on the toroidal magnetic field, the parameter of
$\beta$, and the Mach number, however has outstanding effect on
the density, the gas pressure, the radial infall velocity, and
the angular velocity. The density and the radial thickness of
disk decrease by more advecting of accreting gas that is same at
all part of the disk, the result can be achieved by assuming of
$f$ as a constant amount. Also we see by increasing the amount of
the advection parameter $f$, the gas pressure decreases. By
increasing $f$, the radial infall velocity increases and the
angular velocity decreases. The results are qualitatively
consistent with the results of AF06.

The Mach number profiles in figure 4 imply that the flow of outer
part for all selected amounts of the magnetic field  become super
sonic. We can see this result in polytropic accretion flow by
KF08. The advection parameter decreases the amount of the Mach
number slightly.

The profiles of physical quantities in figure 2 imply that they
have the power of law dependency to $\eta$ in magnetical
domination ($\beta_{in} > 1$). So, by fitting a power function on
data in magnetical domination ($\beta_{in}=10$), we can write
\begin{equation}
    R(\eta) \propto \eta^{-1.66}
\end{equation}
\begin{equation}
    P(\eta) \propto \eta^{-2.58}
\end{equation}
\begin{equation}
   V(\eta) \propto \eta^{-0.01}
\end{equation}
\begin{equation}
    \omega(\eta) \propto \eta^{-1.25}
\end{equation}
\begin{equation}
   B(\eta) \propto \eta^{-0.83}
\end{equation}
\begin{equation}
    \beta(\eta) \propto \eta^{-0.92}
\end{equation}
\begin{equation}
    \mu(\eta) \propto \eta^{0.93}
\end{equation}
\begin{equation}
    \dot{m}(\eta) \propto \eta^{-0.33}.
\end{equation}
The achieved results are different with steady magnetical
dominated accretion flow (Meier 2005, Shadmehri \& Khajenabi
2005).

\section{Summary and Discussion}
In the paper, the equations of time-dependent of advection
dominated accretion flow with a toroidal magnetic field have been
solved by semi-analytical similarity methods. The flow is not
able to radiate efficiency, so we substituted the energy equation
instead of polytropic equation that KF08 had used. A solution was
found for the case $\gamma < 5/3$ that has differential rotation
and viscous dissipation. The flow avoids many of the strictures
of steady self-similar solutions (Narayan \& Yi 1994; AF06;
Ghanbari et al. 2007; Abbassi et al. 2008). Thus, the
radial-dependence of calculated physical quantities in this
approach are different from steady self-similar solution.

Increase of the advection parameter $f$ and the parameter
$\beta_{in}$ will separately increase the infall radial velocity
and decrease the angular velocity. The flow has differential
rotation and is sub-Keplerian in inner part and is
super-Keplerian in large radii in which the behavior is seen in
some astrophysical objects such as M81, M87 and Milky Way (Sofue
1998; Ford \& Tsvetanov 1999).
The solution showed that the flow for all selected amounts of $f$
and $\beta_{in}$ becomes super sonic in large radii and sub-sonic
in small radii that are qualitatively consistent with the results
of KF08.
 The parameter of $\beta$ is a function of position that raises from inner
  to outer and states the magnetic field is more important in large
  radii. It is also consistent with observational evidences in the outer regions
of YSO discs (Greaves et al. 1997; Aitken et al. 1993; Wright et al. 1993)
 and in the Galactic center (Novak et al. 2003; Chuss et al. 2003).

Here, latitudinal dependence of physical quantities is ignored,
while some authors showed that latitudinal dependence is
important in the structure of a disk (Narayan \& Yi 1995; Ghanbari
et. al. 2007). Latitudinal behavior of such disks can be
investigated in other studies. Also we did not consider
relativity effect, If the central object is relativistic, the
gravitational field should be changed. Furthermore, in a
realistic model the advection parameter $f$ is a function of
position and time, other researchers can consider such disks.

\section*{Acknowledgments}
 We wish to thank the anonymous referee for his/her very
 constructive comments which helped us to improve the initial version of the paper; we would
 also like to thank  Wilhelm Kley and Serena Arena for their helpful discussion.

\clearpage
\onecolumn

%#####################################
\input{epsf}
\begin{figure} %[ht]
%\begin{figure*}[!ht]
\begin{center}
%\vspace{-3in}
\centerline
{ 
{\epsfxsize=5cm\epsffile{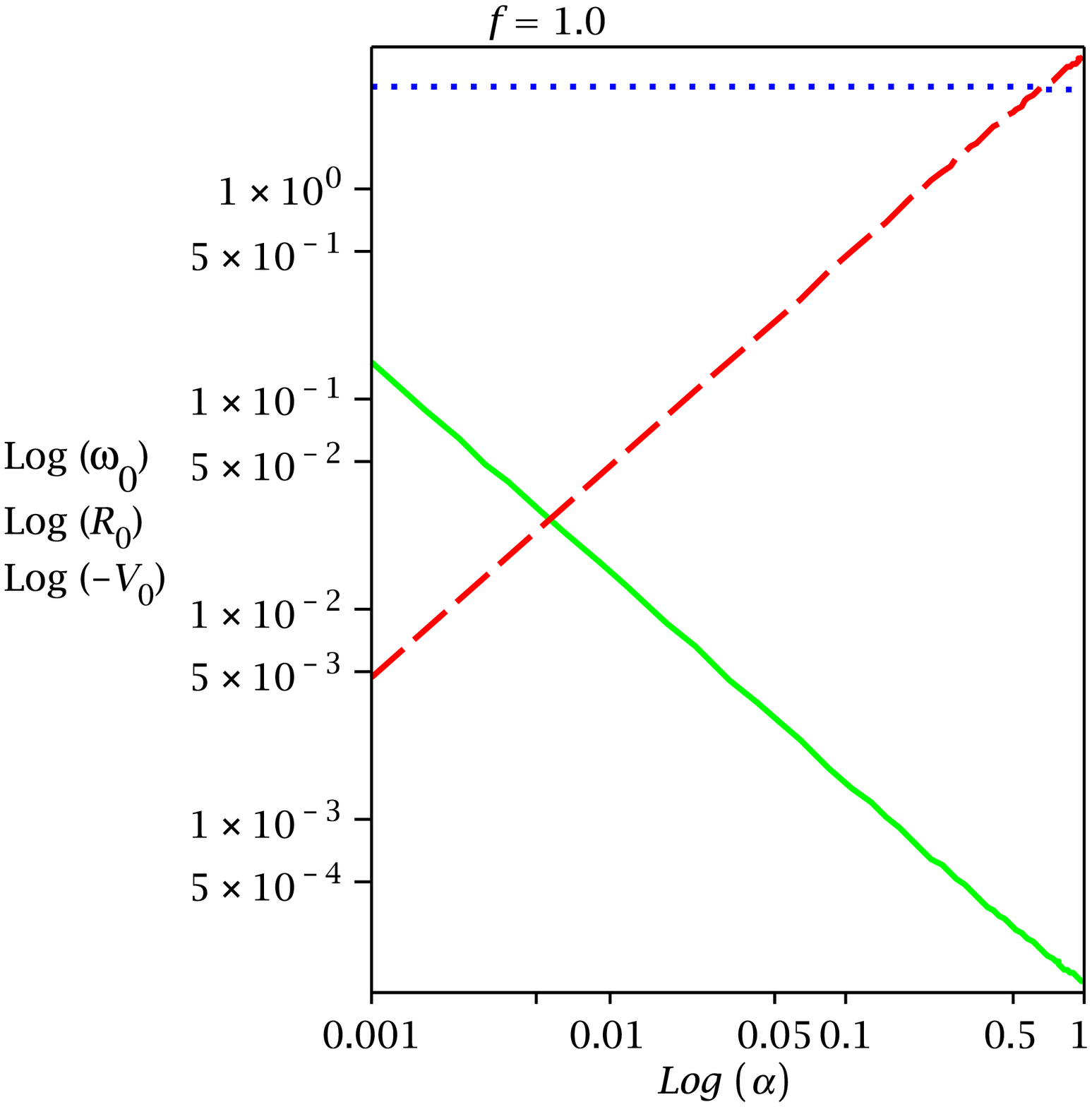}}{\epsfxsize=5cm\epsffile{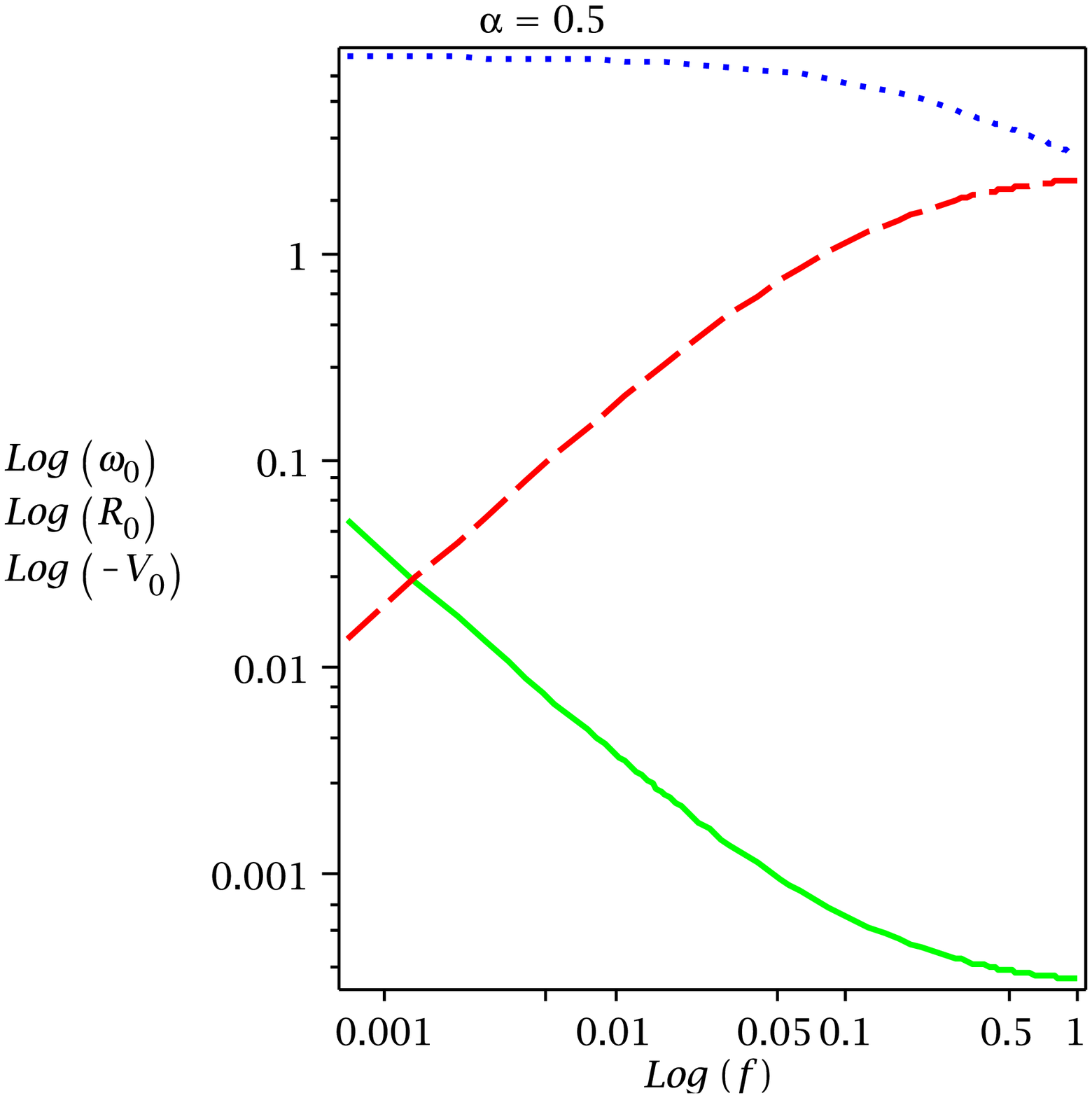}  }
} 
\end{center}
\begin{center}
\caption{Numerical coefficient $\omega_0$ (dotted lines), $R_0$ (solid lines) and $V_0$ (dashed lines)
 as functions of advection parameter $f$ or the the viscous parameter $\alpha$. The ratio of specific heats is set to be $\gamma=1.5$ and
the inner mass accretion rate is $\dot{m}_{in}=0.001$.}
\end{center}
\end{figure}
%#####################################

%#####################################
\input{epsf}
\begin{figure} %[ht]
%\begin{figure*}[!ht]
\begin{center}
%\vspace{-3in}
\centerline
{ 
{\epsfxsize=6cm\epsffile{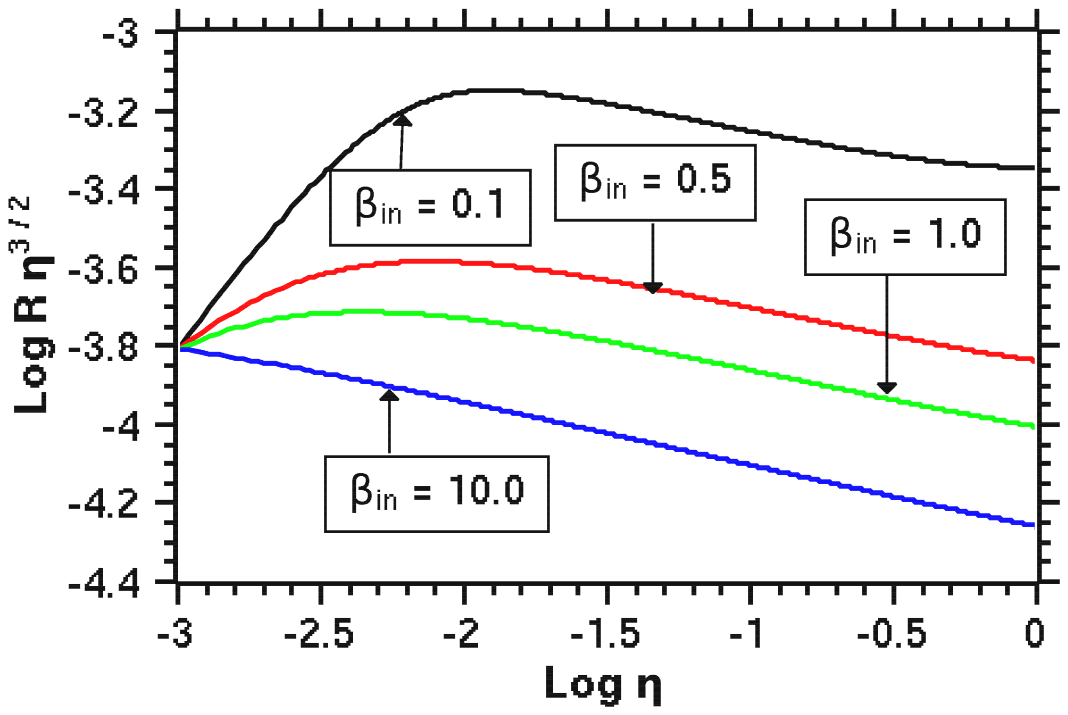}}{\epsfxsize=6cm\epsffile{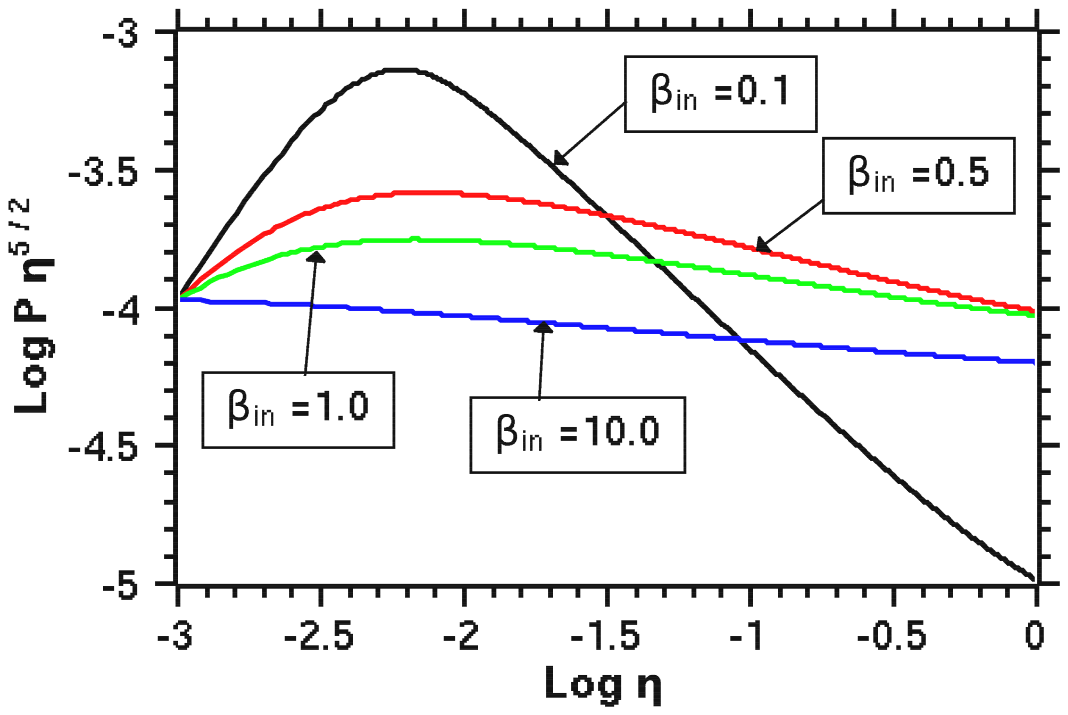}  }
} 
\centerline
{ 
{\epsfxsize=6cm\epsffile{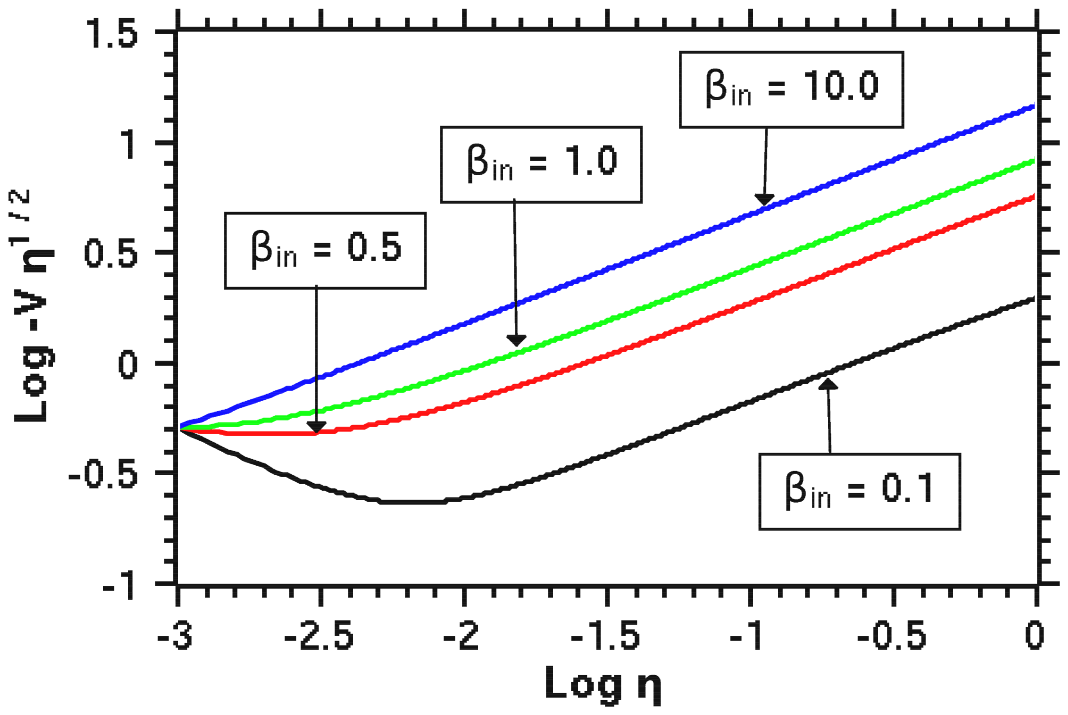}}{\epsfxsize=6cm\epsffile{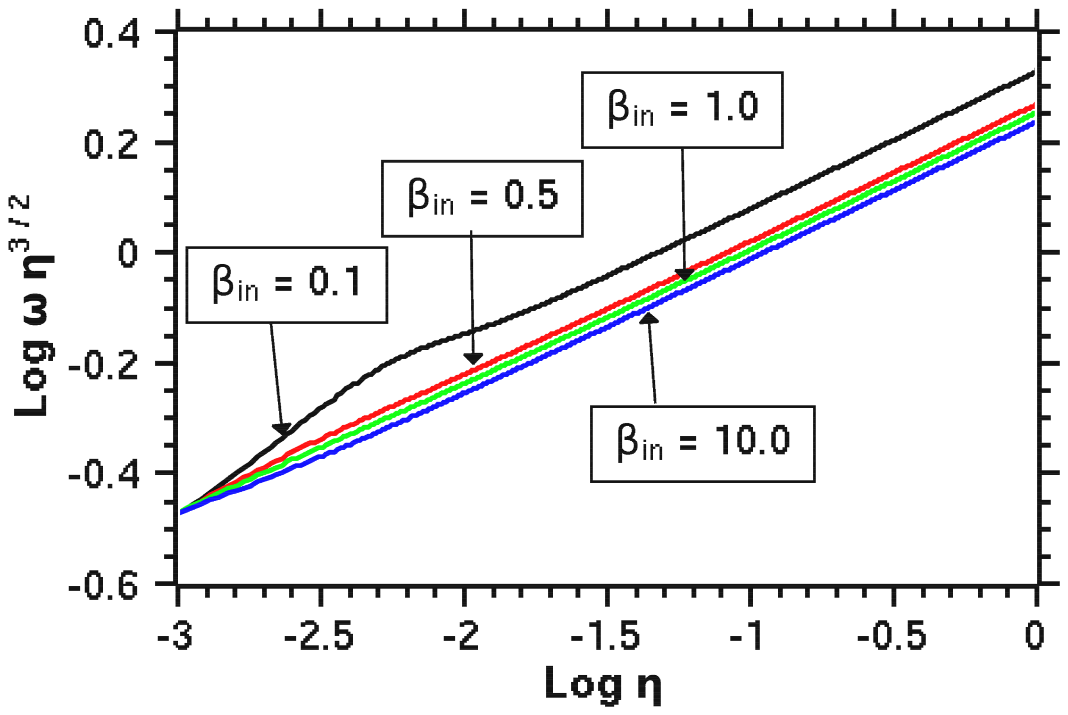}  }
} 
\centerline
{ 
{\epsfxsize=6cm\epsffile{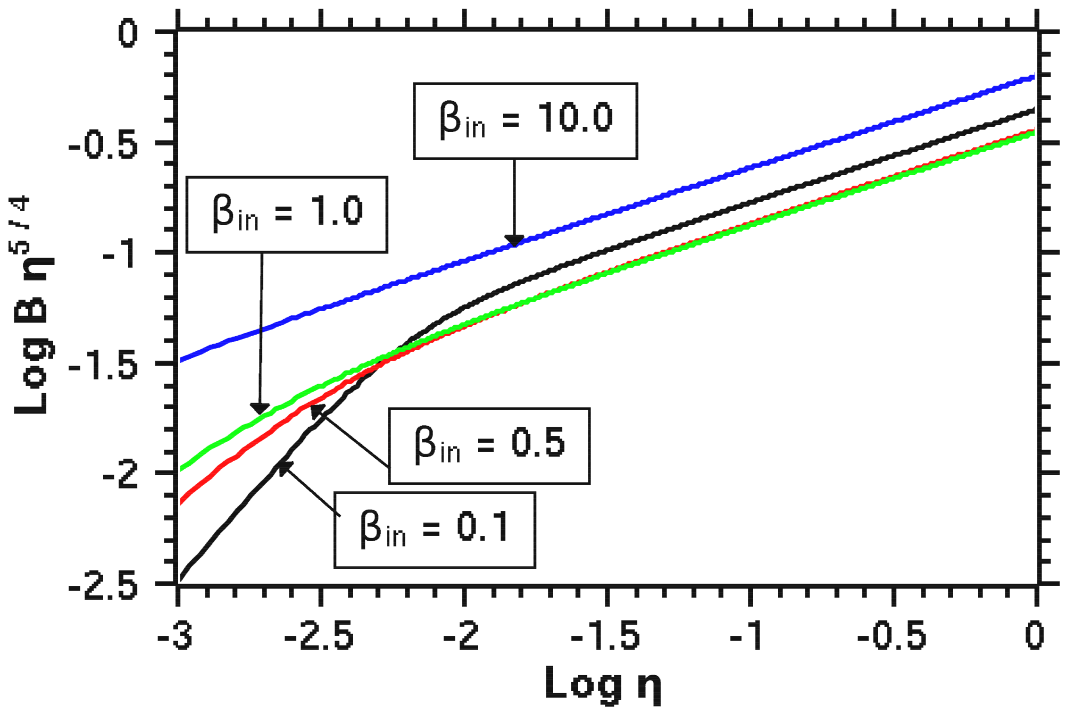}}{\epsfxsize=6cm\epsffile{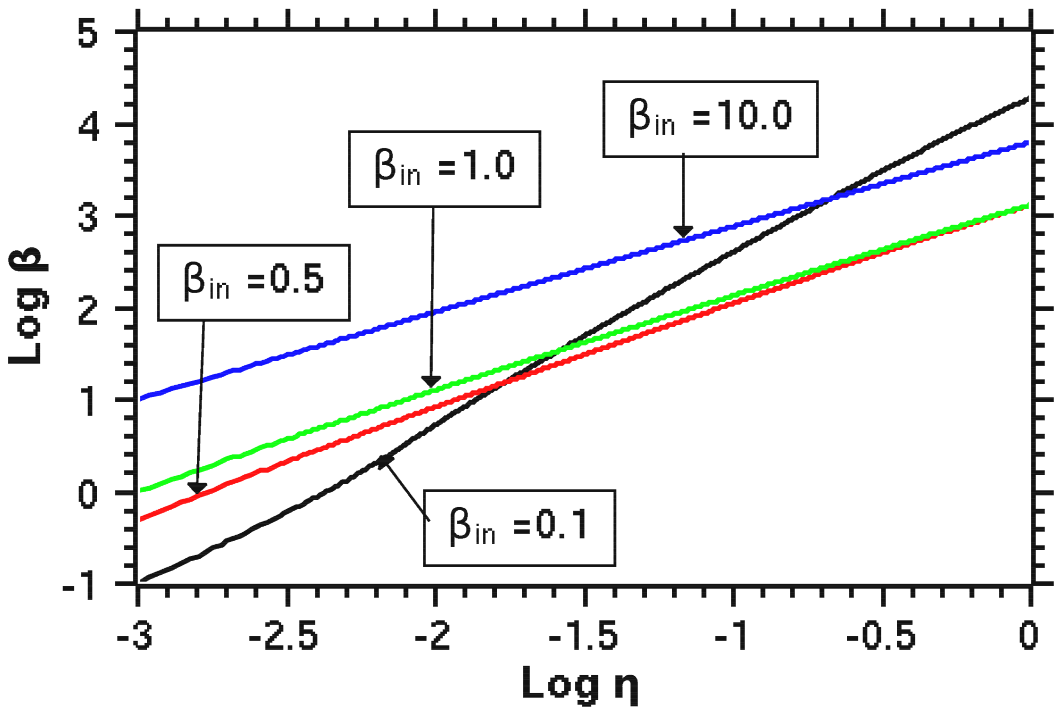}  }
} 
\end{center}
\begin{center}
\caption{Time-dependent self-similar solution for
$\gamma = 1.5$, $\alpha = 0.5$, $f=1.0$, and $\dot{m}_{in}=0.001$.
The lines represent $\beta_{in}= 0.1, 0.5, 1.0, 10$ that $\beta_{in}$ is value of $\beta$ in $\eta_{in}$.}
\end{center}
\end{figure}
%#####################################

%#####################################
\input{epsf}
\begin{figure} %[ht]
%\begin{figure*}[!ht]
\begin{center}
%\vspace{-3in}
\centerline
{ 
{\epsfxsize=6cm\epsffile{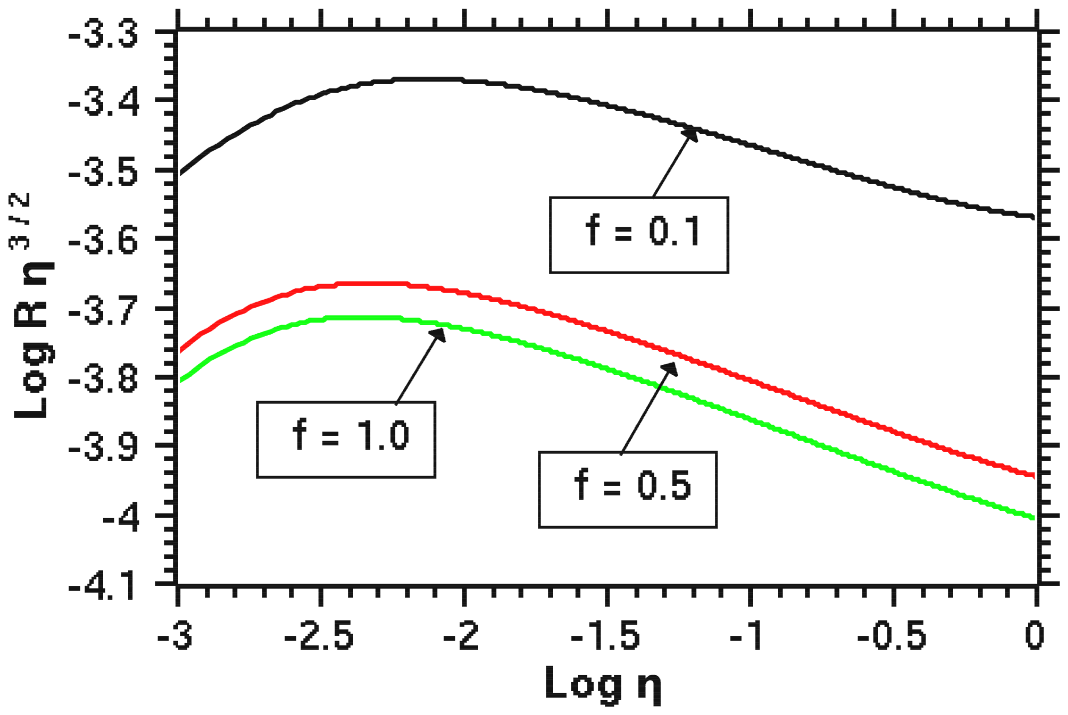}}{\epsfxsize=6cm\epsffile{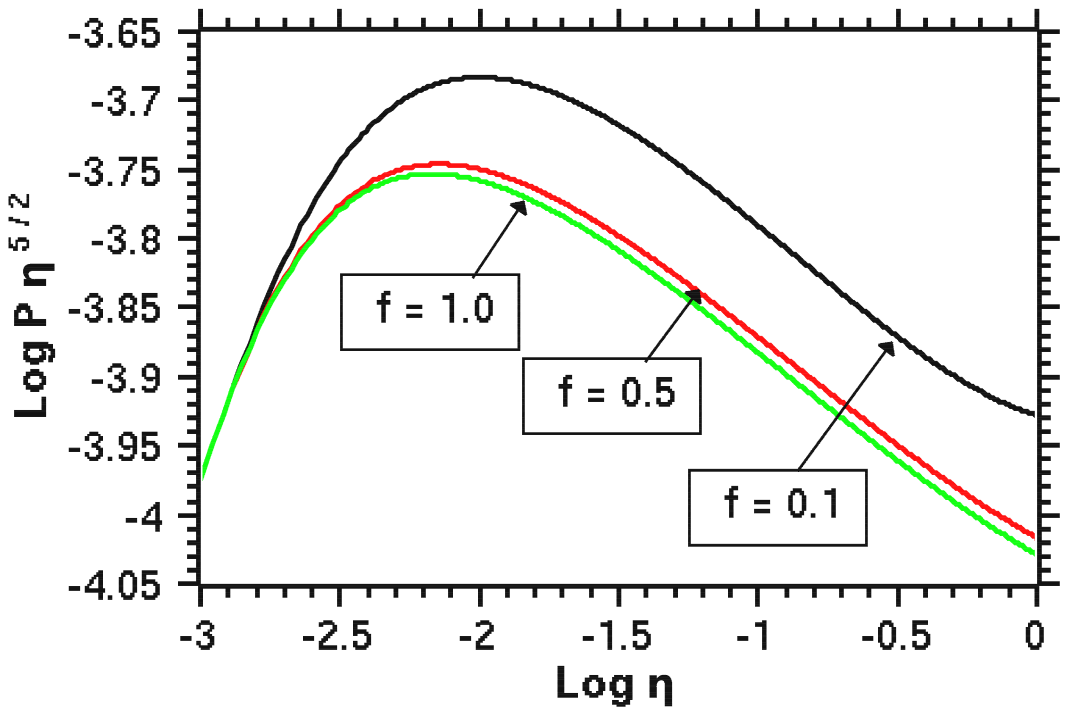}  }
} 
\centerline
{ 
{\epsfxsize=6cm\epsffile{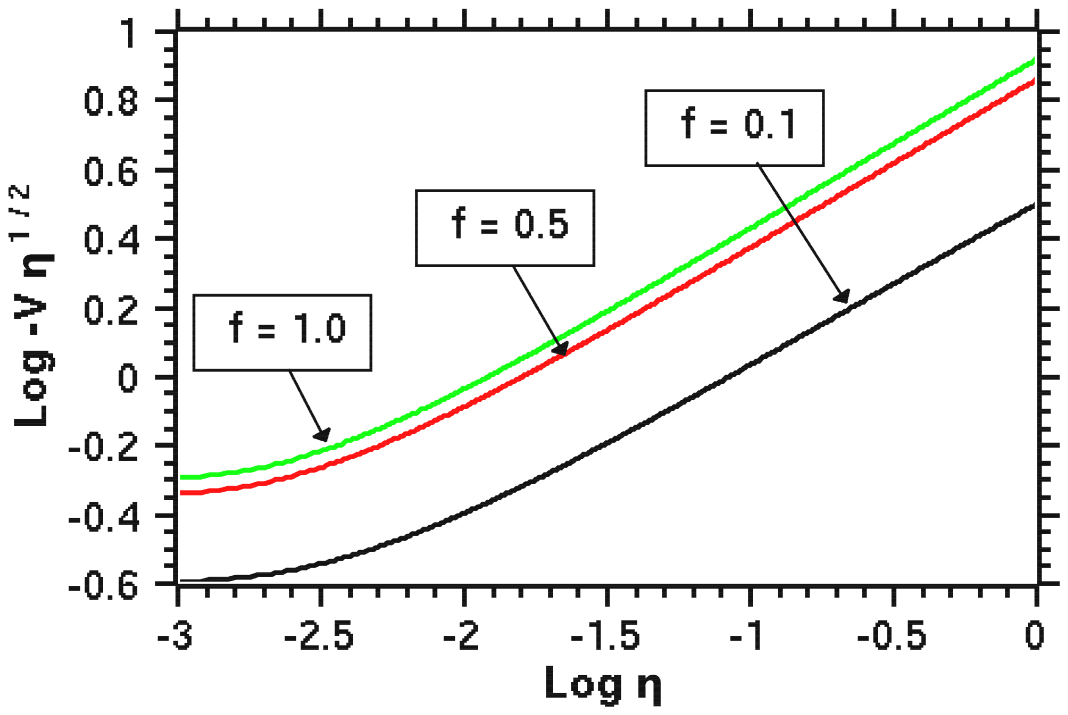}}{\epsfxsize=6cm\epsffile{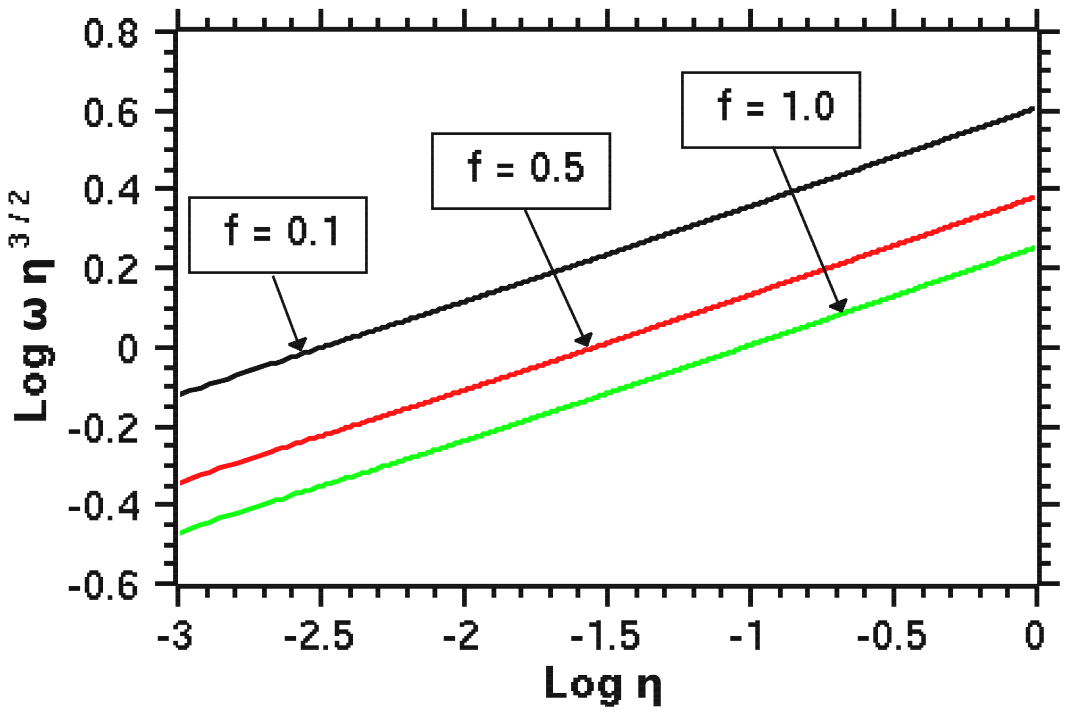}  }
} 
\centerline
{ 
{\epsfxsize=6cm\epsffile{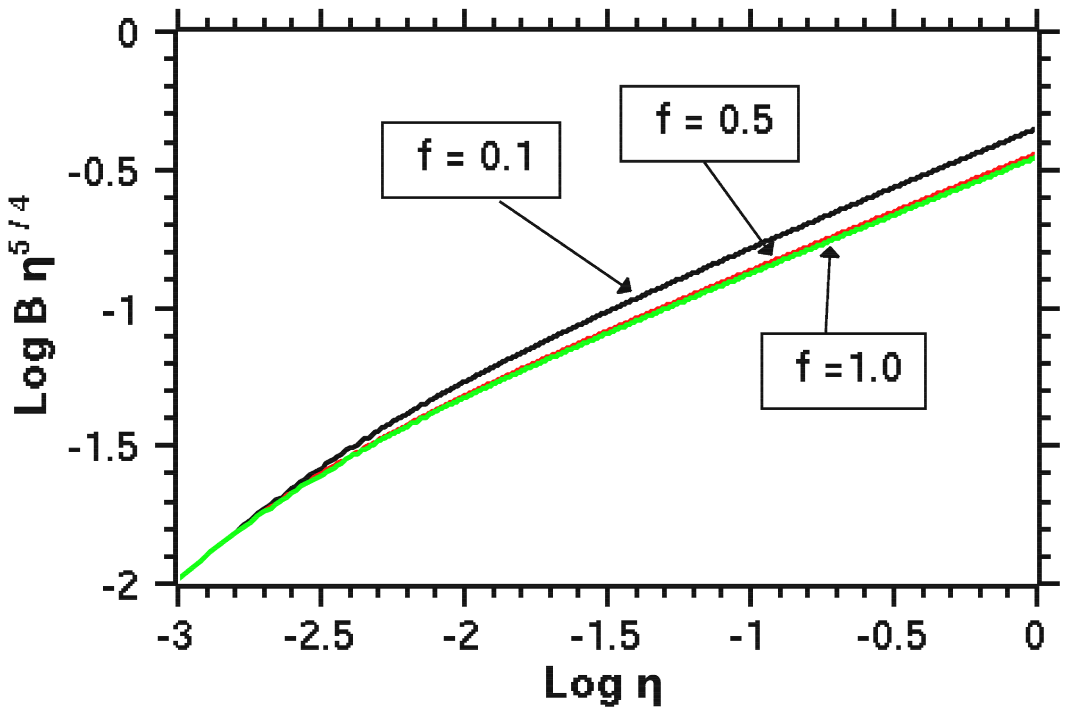}}{\epsfxsize=6cm\epsffile{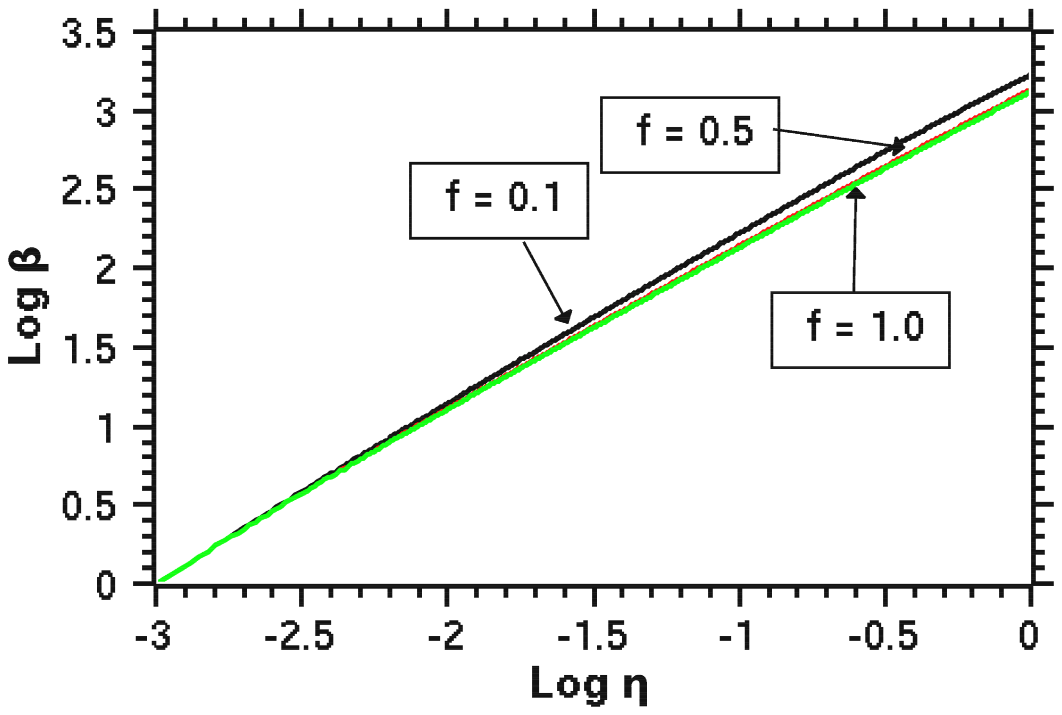}  }
} 
\end{center}
\begin{center}
\caption{Time-dependent self-similar solution for $\gamma = 1.5$, $\alpha = 0.5$, $\beta_{in}=1.0$, and $\dot{m}_{in}=0.001$.
lines represent $f = 0.1, 0.5, 1.0$.}
\end{center}
\end{figure}
%#####################################

%#####################################
\input{epsf}
\begin{figure} %[ht]
%\begin{figure*}[!ht]
\begin{center}
%\vspace{-3in}
\centerline
{ 
{\epsfxsize=6cm\epsffile{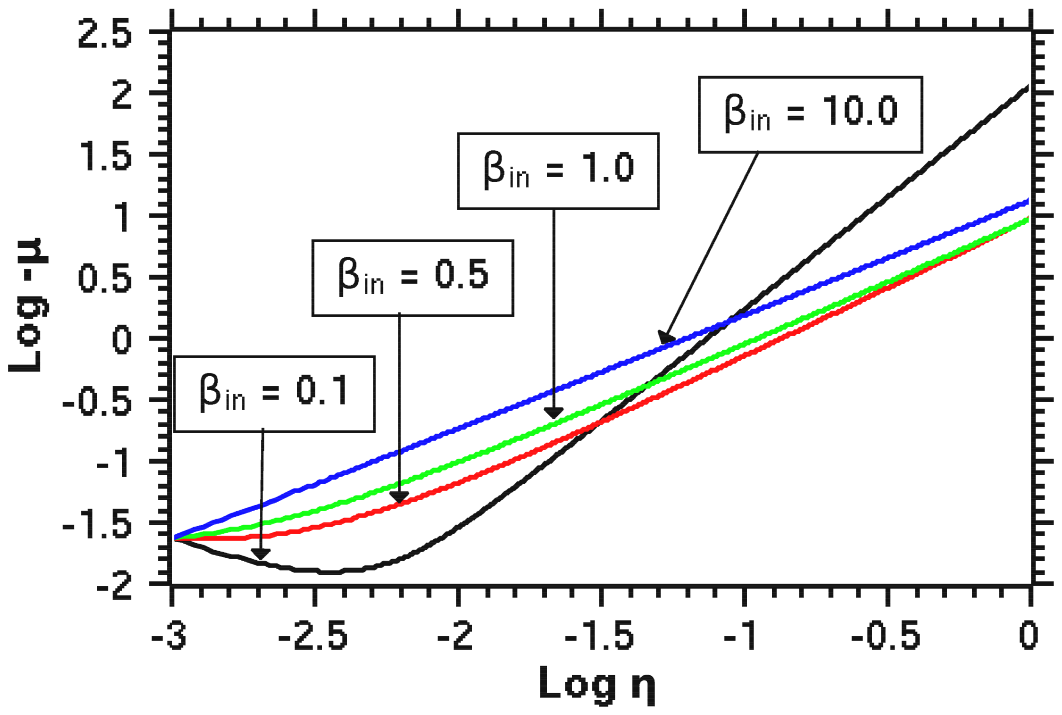}}{\epsfxsize=6cm\epsffile{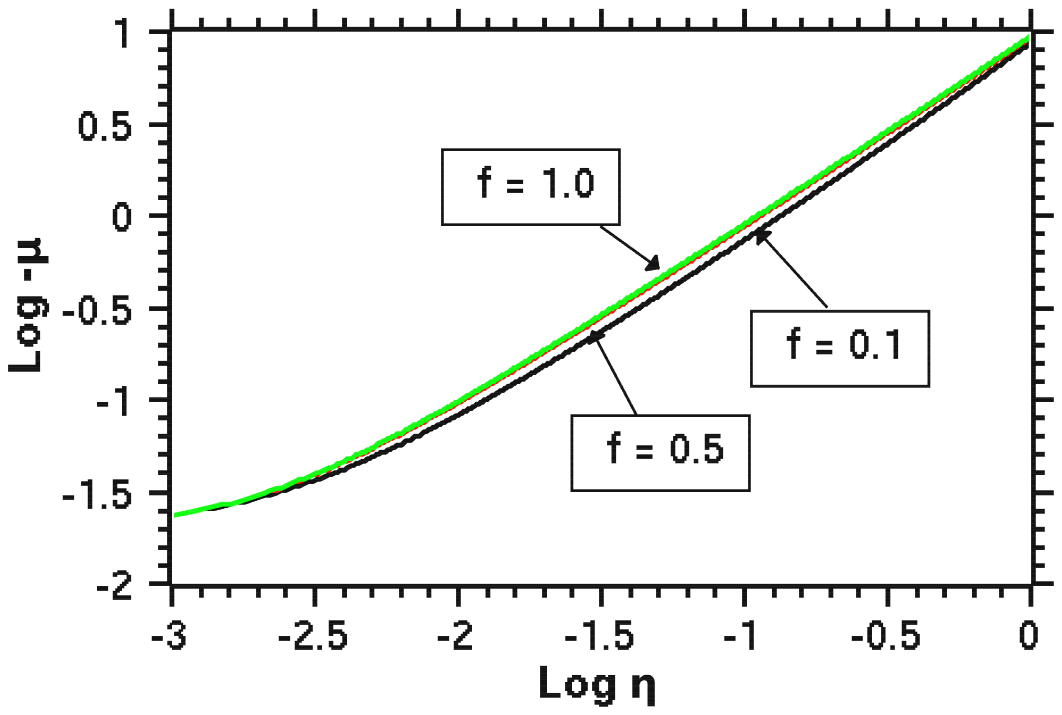}  }
} 
\end{center}
\begin{center}
\caption{Left panel: Mach number profiles for $\gamma = 1.5$, $\alpha = 0.5$, $f=1.0$, and $\dot{m}_{in}=0.001$. Right panel:
Mach number profiles for  $\gamma = 1.5$, $\alpha = 0.5$, $\beta_{in}=1.0$, and $\dot{m}_{in}=0.001$.}
\end{center}
\end{figure}
%#####################################

%\bsp

%\label{lastpage}

\end{document}